\documentclass[twocolumn]{aastex631}

\begin{document}
\title{ High Resolution ($<100$\,pc) Ionization Structures in the ISM of AGN ESO~137-G34}

\author[0000-0002-3626-5831]{D.~{\L}.~Kr\'{o}l}
\affiliation{Center for Astrophysics $|$  Harvard \& Smithsonian, 60 Garden Street, Cambridge, MA 02138, USA}
\affiliation{Astronomical Observatory of the Jagiellonian University, Orla 171, 30-244 Krak\'{o}w, Poland}
\author[0000-0002-3554-3318]{G.~Fabbiano}
\affiliation{Center for Astrophysics $|$  Harvard \& Smithsonian, 60 Garden Street, Cambridge, MA 02138, USA}
\author[0000-0001-5060-1398]{M.~Elvis}
\affiliation{Center for Astrophysics $|$  Harvard \& Smithsonian, 60 Garden Street, Cambridge, MA 02138, USA}
\author[0000-0001-8112-3464]{A.~Trindade~Falcão}
\affiliation{Center for Astrophysics $|$  Harvard \& Smithsonian, 60 Garden Street, Cambridge, MA 02138, USA}
\affiliation{ NASA-Goddard Space Flight Center, Code 662, Greenbelt, MD 20771, USA}
\author[0000-0002-1333-147X]{P.Zhu}
\affiliation{Center for Astrophysics $|$  Harvard \& Smithsonian, 60 Garden Street, Cambridge, MA 02138, USA}
\author[0000-0001-8152-3943]{L.J.Kewley}
\affiliation{Center for Astrophysics $|$  Harvard \& Smithsonian, 60 Garden Street, Cambridge, MA 02138, USA}
\author[0000-0001-9815-9092]{R.~Middei}
\affiliation{Center for Astrophysics $|$  Harvard \& Smithsonian, 60 Garden Street, Cambridge, MA 02138, USA}
\author[0000-0002-0001-3587]{D.~Rosario}
\affiliation{School of Mathematics, Statistics and Physics, Newcastle University, Newcastle upon Tyne NE1 7RU, UK}
\author[0000-0003-4949-7217]{R.~Davies}
\affiliation{Max Planck Institute for Extraterrestrial Physics, Giessenbachstrasse, 85741 Garching bei München, Germany}
\author[0000-0002-2125-4670]{T.~Shimizu}
\affiliation{Max Planck Institute for Extraterrestrial Physics, Giessenbachstrasse, 85741 Garching bei München, Germany}
\author{D.~Hill}
\affiliation{School of Mathematics, Statistics and Physics, Newcastle University, Newcastle upon Tyne NE1 7RU, UK}

\begin{abstract}
We present a detailed, spatially resolved {  Veilleux \&  Osterbrock (VO) diagram} analysis of ESO\,137-G034, a Compton-thick Active Galactic Nucleus (AGN), based on narrowband {\it HST} imaging in [O~III], [S~II], H$\alpha$, and H$\beta$ emission lines. The narrowband optical emission exhibits a bi-conical morphology and traces the diffuse X-ray emission observed with \textit{Chandra}. We dissect the Interstellar Medium (ISM) into Seyfert-, LINER-, and H~II-dominated regions with a resolution of $\sim0.1^{\prime\prime}$ ($\sim20$\,pc at $z=0.009$). To { visualize} the fine spatial structure of the { ISM excitation}, we introduce a new parameter: the Seyfert/LINER Index (SLI), defined as the perpendicular distance of each point in the { VO} diagram from the Seyfert/LINER division line. { SLI mapping reveals fine spatial structures in the ISM excitation especially in the Seyfert ionization bi-cones where the statistical significance of our data is higher.}  Most of the emission within the bi-cones is Seyfert-like, with SLI values consistent with regions of the { VO} diagram that can be modeled with AGN photoionization. In the North-West Seyfert cone three radial SLI peaks suggest episodic nuclear outbursts, with estimated timescales of a few $10^3$\,yrs and $<100$\,yrs duration. The South-East cone shows inhomogeneous excitation, with a higher SLI region in its inner part, coinciding with a wider region of enhanced soft  X-ray emission at the radio lobe inner edge. This region is characterized by line ratios consistent with fast shocks ($>1000$\,km\,s$^{-1}$) induced excitation. The North-West cone is surrounded by a smooth LINER ``cocoon''. The LINER region in the SE cone appears more irregular, with inter-cone LINER points likely shaped by both local ISM structure and shocks. Our findings highlight the complex interplay of different AGN feedback mechanisms in the ISM.
\end{abstract}


\section{Introduction} \label{sec:intro}

The effect of an active galactic nucleus (AGN) on its host galaxy may be crucial for galaxy evolution \citep[][]{Heckman14}. The connection between an AGN and its host is  { evident from the tight correlation} between the mass of the central supermassive black hole (SMBH) and the stellar content of the host galaxy \citep[e.g., $M-\sigma$ relation][]{Kormendy13}. AGNs are also a {powerful source of energy \citep{Fabian12}, potentially driving the feedback processes that  could resolve the long-standing problem of} the discrepancy between the observed galaxy luminosity function and the galaxy mass function predicted by $\Lambda$CDM simulations of large-scale structure, a long-standing issue in galaxy evolution theory \citep{Schechter76,Benson03}. Recent results suggest that the SMBH-to-galaxy mass ratios at high redshifts are extremely high, indicating very rapid initial black hole growth and/or a heavy seed origin \citep[see, e.g.,][]{Adamo24, Bogdan24, Maiolino24}. 

\begin{figure}[thp!]
    \centering
    \includegraphics[width=0.49\textwidth]{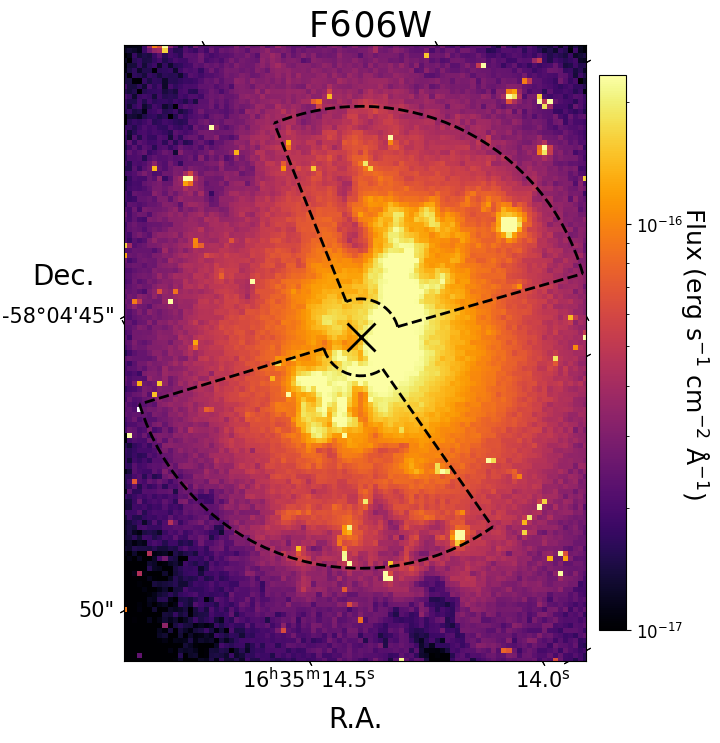}
 \caption{HST/F606W image of ESO~137-G034 \citep[][]{Malkan98}. Dashed black lines mark the ionization cones, as defined by the diffuse X-ray emission (Paper~I). The black ``X'' marks the position of the nucleus, defined as the centroid of the [O~III] continuum emission. }
    \label{fig:F606}
\end{figure}
High spatial resolution observations across the electromagnetic spectrum enable us to probe AGN feedback { in action} and investigate the underlying physical mechanisms, including interstellar medium (ISM) photoionization, jet–ISM shocks, and large-scale winds \citep[][]{Fabbiano22rev}. Diagnostic diagrams comparing ratios of optical and UV emission lines, e.g. the BPT-style (Baldwin–Phillips–Terlevich) diagrams, are powerful tools for distinguishing between star-forming, AGN-dominated, and low-ionization nuclear emission-line region (LINER) dominated galaxies. { The classifications are based on a comparison of the [O~III]/H$\beta$ and [N~II]/H$\alpha$ (N-BPT), [O~I]/H$\alpha$ (O-BPT) \citep[][]{Baldwin81,Kewley06}, or [S~II]/H$\alpha$ \citep[VO diagrams,][]{Veilleux87} emission-line ratios.}  The spatially resolved version of this {diagnostic}, known as { BPT, or VO mapping}, allows the isolation of regions dominated by different ionization processes. The VO map of the nearby Compton Thick (CT) Seyfert\,2 galaxy NGC~5643\citep[][]{Cresci15}, revealed spatially extended, predominantly Seyfert-like emission surrounded by a thin LINER-like cocoon. With the {\it Hubble Space Telescope}\,(HST), such analyses can be performed at $\sim0.1^{\prime\prime}$ scales, corresponding to the linear scales of  tens of parsecs in the nearest AGNs \citep{Maksym16,Ma21,Maksym21,Trindade25}
In this paper, we investigate spatial variations {in the } {excitation state} of ESO\,137-G034 ISM . To this end,  we have constructed a { VO} map that traces the positions of points on the { VO}  diagram rather than relying on a simple LINER/Seyfert dichotomy, the Seyfert/LINER Index (SLI).  { In particular, we examine the SLI values within the LINER cocoon to assess whether they change systematically with distance from the nucleus. We also investigate regions characterized by distinctly high or low SLI values through excitation modeling and comparison with X-ray, radio, and infrared emission, in order to determine whether variations in SLI reflect differences in excitation mechanisms and underlying ionization source.  }

ESO\,137-G034 is an S0/a galaxy hosting a Seyfert type II AGN \citep[Fig.\,\ref{fig:F606};][]{Malkan98,Ferruit00} at $z=0.009$ \citep[$D_{L}=41.21$\,Mpc; $1^{\prime\prime}\simeq200\textrm{pc}$,][]{Koss22} in which extended X-ray emission was discovered with {\it Chandra} \citep{Ma20}. In Paper I (Kr\'{o}l et al, 2026 in press), we presented the results from $230$\,ks X-ray follow-up observations with {\it Chandra},  which revealed a {bi-conical structure in the diffuse X-ray emission}, extending to kiloparsec scales. Based on spectral modeling and the morphology of the X-ray emission in different energy bands, we argued that the diffuse emission is a mixture of photoionized gas and shock-heated plasma, with the South-East (SE) cone exhibiting stronger shock signatures and the North-West (NW) cone being dominated by photoionization.

This paper is organized as follows. After the Introduction, we describe the observations, data reduction, and reddening correction (Sec.\,\ref{sec:data_analysis}). In Sec.\,\ref{sec:bpt}, we present the { VO}  mapping and we introduce {the definition of the SLI}. In Sec.\,\ref{sec:discussion}, we interpret our results and compare the {{ VO}  diagnostics} with multi-wavelength data. In Sec.\,\ref{sec:summary}, we summarize our findings.

\section{HST observations and data reduction }\label{sec:data_analysis}
\subsection{Data}

\begin{deluxetable}{ccccc}[t]
\label{tab:observations}
\tablecaption{HST observations}
\tablehead{\colhead{Instrument} & \colhead{Filter}&\colhead{Proposal ID}   &\colhead{$t_E$}  &\colhead{Note}  \\ & \colhead{} & \colhead{} & \colhead{[s]} & \colhead{}}
\startdata
    WFPC2 & FR533N & 6419  & 1600  &  [O~III]$\lambda5077$ \\
    WFPC2& F547M &6419  & 200  &  	[O~III] continuum  \\
    WFPC2 & FR680P15 & 6419  & 1400  &  H$\alpha$+[N~II] \\
    WFPC2& F791W &6419  &  200  &  	H$\alpha$ continuum  \\
    WFC3 & F673N & 16841  & 2000  & [S~II]$\lambda$6716,31\\
    WFC3& F763M & 16841  &  262  &  	[S~II] continuum  \\
    WFC3 & FQ492N & 16841  & 2000  & H$\beta$\\
    WFC3& F547M & 16841  &  254  &  	H$\beta$ continuum  \\
\hline
\enddata
\tablecomments{Data set DOI: \url{https://10.17909/gdjr-yc45}}
\end{deluxetable}

ESO\,137-G034 has been observed multiple times with the {\it HST} (Tab.\,\ref{tab:observations}). 
To construct high-resolution, multi-wavelength maps of the AGN and the VO diagram, we {compiled} data covering the key diagnostic lines: [O~III], $H{\alpha}$, $H{\beta}$, and [S~II]. Observations using the narrow-band filters FR533N and FR680P15, which cover the [O~III] and $H{\alpha}$+[N~II] emission lines, respectively, along with the {wide- and medium-band filters} F791W and F547M to probe the associated continua, were performed with the Wide Field and Planetary Camera 2 (WFPC\,2) instrument \citep{Ferruit00}. {Additional observations using the \textit{Wide Field Camera 3} (WFC\,3)} include narrow-band filters from the F673N and FQ492N, covering the [S~II] and $H{\beta}$ emission lines, respectively, {along with} the F763M {and} F547M filters for the associated continua.

\subsection{Data {Analysis and Reduction}}\label{sec:dust}

\begin{figure*}[thp!]
    \centering
    \includegraphics[width=0.99\textwidth]{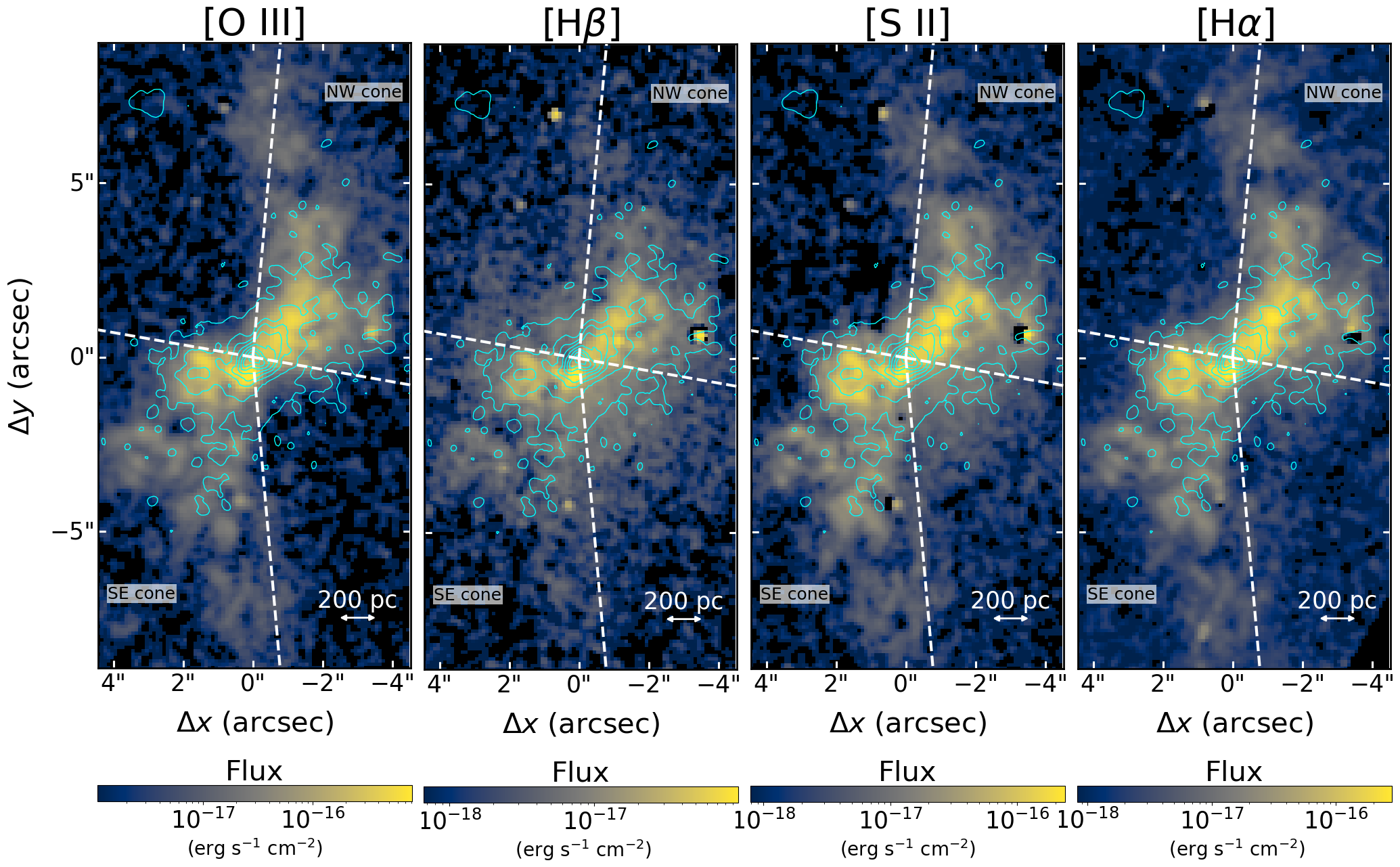}
 \caption{Continuum-subtracted narrow-line emission images of the ESO\,137-G034 in [O~III], H$\beta$, [S~II], and H$\alpha$, from left to right. { Cyan contours trace the extended X-ray emission ($0.3-2.0$~keV), and the cones marked with dashed white lines were obtained based on the azimuthal profile of the diffuse X-ray emission (Paper~I). We aligned the X-ray and [O~III] centroids to define the nucleus. }}
    \label{fig:maps}
\end{figure*}

We began the data analysis by removing cosmic rays from each exposure using the {\tt cosmicray\_lacosmic} algorithm from the {\tt ccdproc} \footnote{Implementation: \url{https://pypi.org/project/astroscrappy/}} package \citep[][]{Dokkum01}. Subsequent {processing steps were carried out} using {\tt DrizzlePac} \citep[][]{Hoffmann21}. Exposures were aligned using {\tt TweakReg}, and final images were drizzled with {\tt AstroDrizzle} to a pixel scale of $0.0996^{\prime\prime}$. We used the {\tt PHOTFLAM} and {\tt PHOTBW} keywords from the headers for flux calibration, converting the images to erg\,s$^{-1}$cm$^{-2}$ units.

We then subtracted the background from both the continuum and line images, determined as the median emission value of line-free regions. To create the continuum-subtracted emission-line {maps}, we subtracted the bandwidth-scaled continuum from the narrow-band filter emission on a pixel-by-pixel basis. When necessary, the scaling was adjusted manually to avoid over- or under-subtraction.

{ The H\,$\alpha$ emission observed through the FR680P15 filter is contaminated by the [N~II]\,$\lambda\lambda$6548,84 lines. Based on spectroscopic data from  Siding Spring Southern Seyfert Spectroscopic Snapshot Survey \citep[S7,][]{Dopita15}, we adopt a nominal value of approximately 45\% for the H\,$\alpha$ contribution within the FR680P15 filter. However, this contribution can vary depending on the location within the galaxy. Data from the S7 indicate that the H\,$\alpha$ fraction ranges between 40\% and 48\% { in the source region}. Since the spatial resolution of S7 ($\sim1^{\prime\prime}$) is significantly lower than that of HST, in Appendix~\ref{sec:app} we explore how VO diagrams change for a broader range of H\,$\alpha$ contributions-namely, 35\% and 55\% as the lower and upper bounds. Although the exact line ratios vary under these assumed contributions, the overall morphology and key conclusions remain unchanged.}

\subsection{Emission-line images }
{Continuum-subtracted emission-line images of ESO~137-G034 are shown in Fig.~\ref{fig:maps}, from left to right: [O~III], H$\beta$, [S~II], and H$\alpha$. Emission in all lines traces a similar bi-conical morphology, elongated in the north–south direction. The smoothed X-ray contours from Paper~I are overlaid on optical maps. The X-ray morphology closely follows the optical emission, particularly in the NW cone. The bi-conical regions, defined based on the azimuthal profile of the diffuse X-ray emission, align well with the direction of the extended optical emission observed in all maps. The diffuse H$\beta$ emission is the weakest and does not extend beyond $\sim800$~pc. The [S~II], H$\alpha$, and [O~III] emissions show a strong excess along the direction of the dust lane (see red regions in Fig.~\ref{fig:dust}) beyond this radius. Knots of enhanced emission are present in the inner part of the SE cone and are most pronounced in [O~III] and H$\alpha$.}
The X-ray and optical images were aligned under the assumption that the nucleus position in the X-ray (taking the centroid of the emission in the $5.5$–$7.0$\,keV band) and the nucleus position in the optical emission are coincident. The centroid around the maximum of the [O~III] continuum emission was assumed to represent the nucleus position in the {\it HST} data. { All  shifts are within the {\it Chandra } and {\it HST} astrometric uncertainties. }

\section{{ VO}  mapping}\label{sec:bpt}

\begin{figure*}[thp!]
    \centering
    \includegraphics[width=0.95\textwidth]{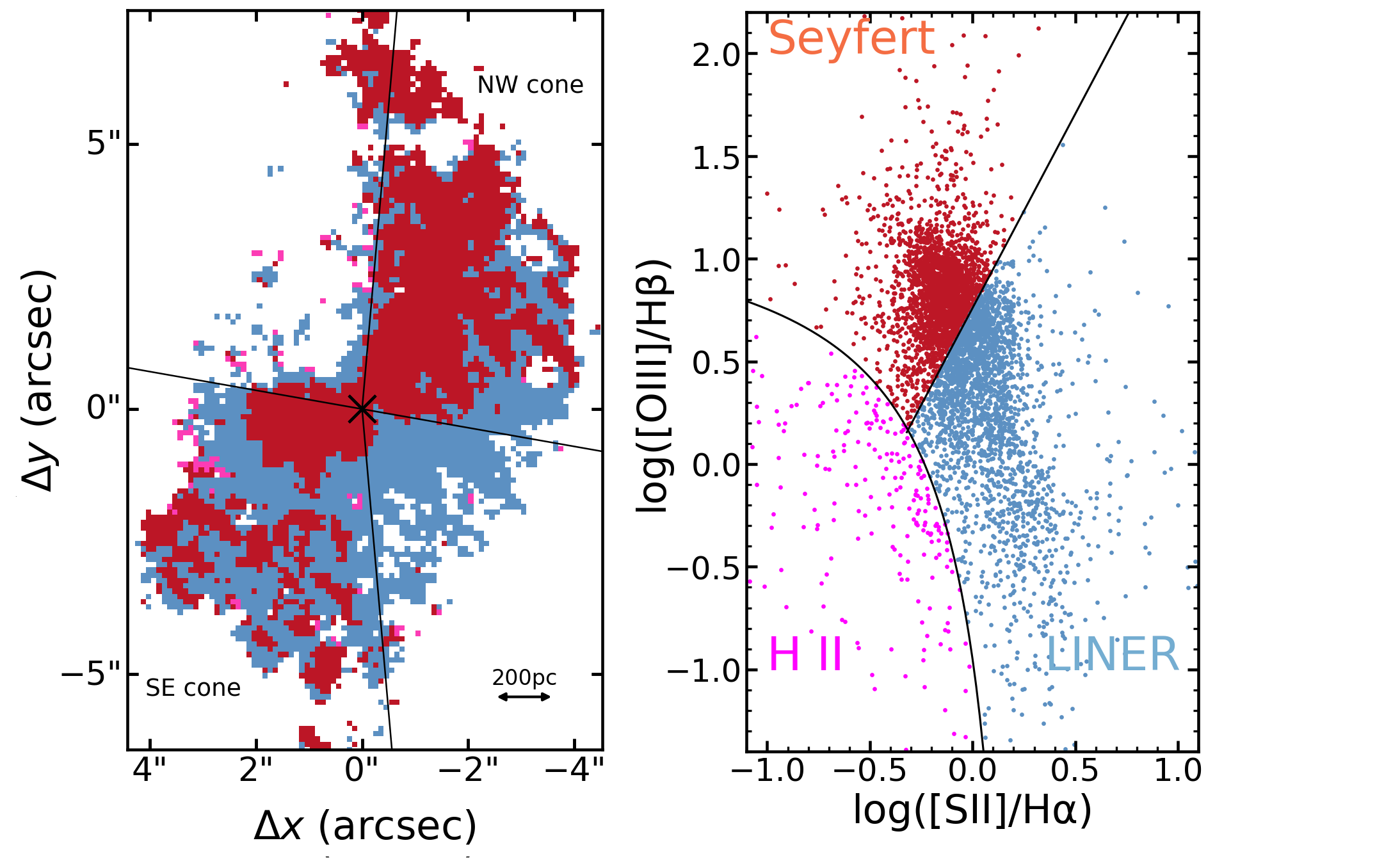}
 \caption{ Left panel: {  { VO}  map of ESO~137-G034.  Colors indicate the ionization classification based on line ratios}: red for Seyfert-like points, blue for LINER-like points, and magenta for H~II region-like points. The dashed outlines mark the ionization bicone defined by the extended X-ray emission (see Paper~I for details).  The black ``X'' marks position of the nucleus. Right panel: {corresponding { VO}  diagram, with the same color scheme representing the classification of individual pixels.}}
    \label{fig:bpt}
\end{figure*}

We constructed the { VO }excitation map for ESO\,137-G034 by comparing the values of $\log$([O~III]/H$\beta$) and $\log$([S~II]/H$\alpha$) \citep[][]{Kewley06}. Fluxes were calculated on a pixel-by-pixel basis, only for pixels with non-background-subtracted narrow-band filter fluxes at or above the $3\sigma$ level, where $\sigma$ is the standard deviation of pixel count in a background-dominated region \citep[][]{Ma21}.

In Fig.\,\ref{fig:bpt}, we present the spatially resolved { VO}  map and {corresponding diagnostic diagram}. In the left panel, pixels {classified as Seyfert-like are shown in red, LINER-like regions in blue, and H~II-like regions in magenta. The corresponding { VO}  diagram is shown in the right panel. }

Most of the extended emission is characterized by Seyfert-type excitation (red). The northern part of the source exhibits a clear, cone-like Seyfert-type region surrounded by a LINER cocoon morphology (blue). {In contrast, } the southern part shows a more complex structure, with LINER-type points mingled with the Seyfert emission.  Small pockets of H~II like emission show up mostly outside the LINER cocoon in the NE cross-cone.

\subsection{Seyfert-LINER Index}
\begin{figure*}[]
    \centering
    \includegraphics[width=0.95\textwidth]{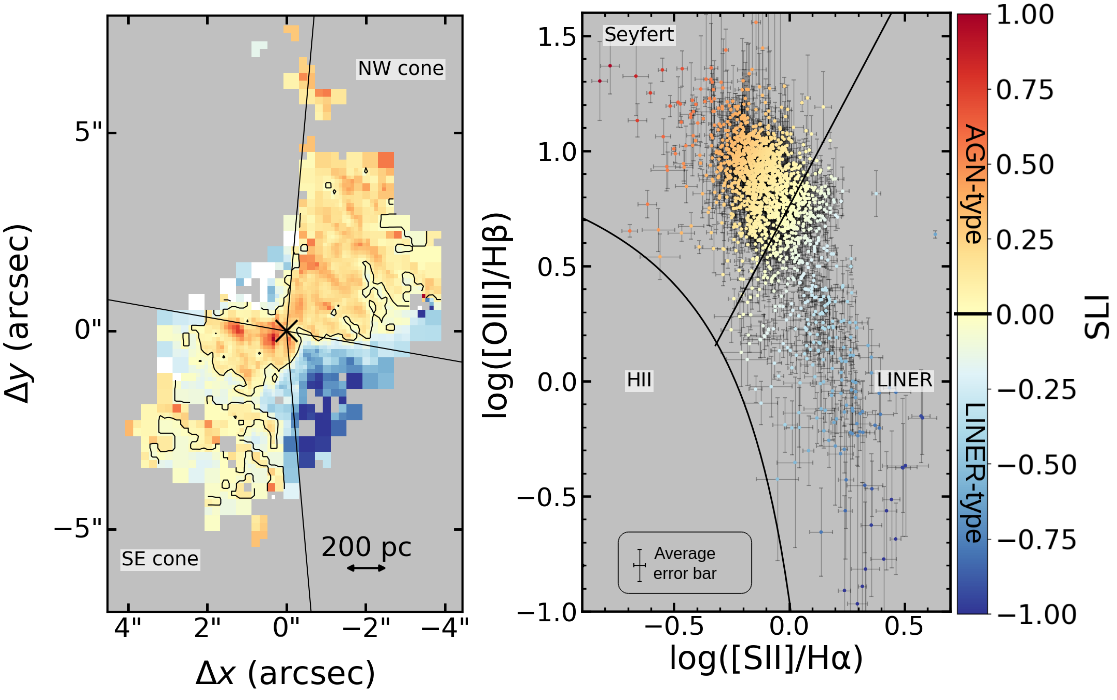}
     \caption{ Left panel: { VO}  {map} showing Seyfert- and LINER-type points, color-coded {by value of SLI, which quantifies the distance from the Seyfert/LINER division line in the { VO}  diagram.} Black contours trace points on the Seyfert/LINER division line: SLI$=0$.  The black ``X'' marks the position of the nucleus. { Data have been adaptively binned to reach S/N$=9$ in each filter, see the text for details.} Right panel:{ Corresponding { VO}  diagram, with red indicating Seyfert-like excitation, and blue denoting LINER-like excitation. For Seyfert and LINER points, the color intensity scales with the SLI value.}}
    \label{fig:bpt_sli}
\end{figure*}
To { better visualize} the spatial distribution of different excitation populations of LINER- and Seyfert-type points, we define the Seyfert–LINER Index (SLI) as the {perpendicular} distance of each point from the Seyfert/LINER division line on the { VO}  diagram. We color-code SLI values on both the { VO}  map and diagram (Fig.\,\ref{fig:bpt_sli}). {By construction,} LINER points have SLI $<0$ and Seyfert points SLI $>0$. The black contour in the left panel represents points that lie on the Seyfert/LINER division line.

{To account for the uncertainties in the SLI values, we adopted the standard deviation of the pixel counts in a background-dominated region, $\sigma$, as the uncertainty for each individual filter observation \citep[][]{Ma21} and propagate the errors.  For the details see Appendix~\ref{app:errors}}. { We adaptively binned the data \citep[][]{Li23} in all narrow- and wide-band filters to achieve a target signal-to-noise ratio of 9. As a result, the average uncertainties in the $\log$([O~III]/$H\beta$) and $\log$([S~II]/$H\alpha$) line ratios were reduced to $\sim0.1$ dex and $\sim0.05$ dex, respectively. }

{We find no clear correlation} between a pixel's SLI value and its distance from the nucleus defined as the centroid of the [O~III] continuum emission, marked with ``X'' in the left panel of both Fig.\ref{fig:bpt} and \ref{fig:bpt_sli}. 

In the SE cone, the SLI map reveals inhomogeneities unseen in the traditional { VO}  map, showing regions with high SLI values (up to $\sim0.7$, heavily Seyfert dominated) at $\le80$\,pc from the nucleus. { The highest SLI values ($>0.45$) are concentrated in two regions along the northern boundary of the SE cone, at projected distances of $\sim100$ pc and $\sim300$ pc from the nucleus.} Beyond $\sim200$\,pc, the interior of the SE cone is { partially} classified as LINER-type with SLI values $\gtrsim-0.15$, {indicating proximity to the Seyfert–LINER boundary.} The NW cone is characterized by the moderate SLI  values in the range of $\sim0.1-0.4$. It is surrounded by a thin LINER cocoon ($\sim100-200$\,pc) with SLI values between $\sim-0.3$ and $0$. LINER-type emission in the cross-cone regions exhibit lower SLI values, generally SLI $<-0.2$.

{The { VO } diagrams are generally not sensitive to reddening, as the emission lines used to compute the ratios are close in wavelength. However, in the case of narrow-band imaging with { {\it HST}}, extinction can be more impactful, as the central wavelengths of the wide filters necessarily used for continuum extraction differ more from the narrow-line filters. In Appendix~\ref{sec:app2}, we examine how dust affects the resulting { VO}  and SLI maps. The reddening corrections influence the SLI values only in regions located within the dust lane (red area in Fig.~\ref{fig:dust}) {, where SLI is modified by $\sim0.2$. Apart from the dust lane SLI changes by $<0.05$.}. In most regions, the extinction correction increases the uncertainties without changing the inferred excitation classification. In the main  body of the paper we present the figures based on the raw data, unless noted otherwise.}

\subsection{Seyfert-LINER Index and excitation models}

\begin{figure*}[thp!]
    \centering
    \includegraphics[width=0.99\textwidth]{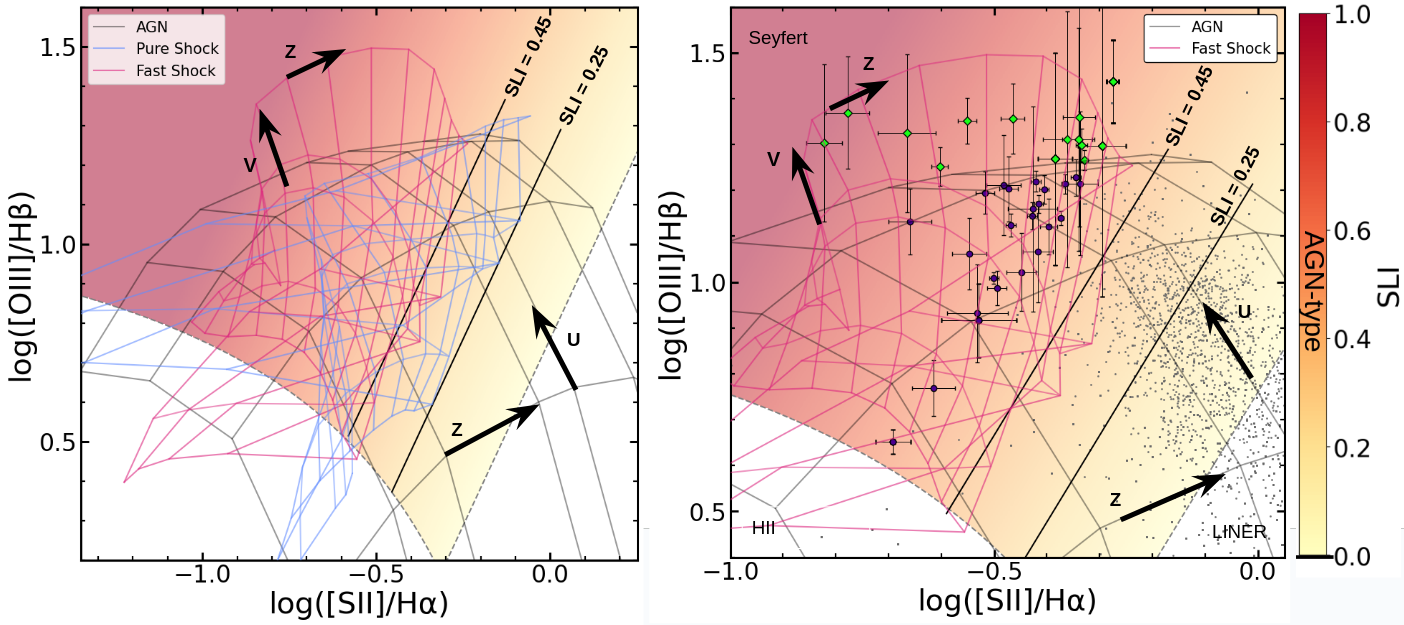}
 \caption{{ Left panel: comparison of the  shocks and AGN excitation models with SLI values. The background color scale indicates the SLI value, with three black lines denoting SLI$=0.25$ and $0.45$.  AGN models with gas pressure of $\log{(P/k)}=7.4$ and $\log (E_{\text{peak}}/\text{keV})=-1.0$ are shown with black grids that consist of constant metallicity lines, Z (from left to right) $12+\log(\rm O/H)=8.43$, $8.70$, $8.80$, $9.02$, $9.26$ and constant ionization parameter lines $\log(\rm U)$ (from bottom to top) $-3.8$, $-3.4$, $-3.0$, $-2.6$, $-2.2$. The``fast shock'' models with $\log{(P/k)}=10.2$ and $\eta_M=0.001$ are shown with red grids with constant metallicity lines ($12+\log(\rm O/H)=8.43$, $8.70$, $8.80$, $9.02$, $9.26$) and constant shock velocity lines ($V_s=328$, $370$, $418$, $472$, $533$, $601$, $679$, $766$, $865$, $976$, $1102$, $1244$, $1404$, $1585$\,kms$^{-1}$).  The ``pure'' shock models are plotted with blue grids for the same $12+\log(\rm O/H)$ and $V_s$ values. Right panel:  ESO 137-G034 { VO } diagram, with over-plotted the fast shock and AGN photoionization model grids. Colored points correspond to pixels with SLI $>0.45$: the purple points can be explained by either AGN photoionization or fast-shock excitation, whereas the green diamonds are inconsistent with AGN photoionization.  In both panels, dashed black lines delimit the Seyfert, LINER and HII type emission regions. Arrows indicate the direction of increase metallicity (Z) and $V_s$ shock velocity(v)  for shock models and photoionization (U) for AGN models. }}
    \label{fig:emodel}
\end{figure*}

{ To investigate the link between the SLI and the excitation mechanism, }{we compare radiative shock models and AGN photoionization models to the observed ratios of emission-line fluxes on the { VO}  diagram (Fig.~\ref{fig:emodel}). We adopt the isobaric, radiation-pressure–dominated AGN photoionization models of \citet{Zhu23}. The incident AGN spectral energy distribution (SED) is generated with OXAF \citep{Thomas16}, a simplified implementation of OPTXAGNF \citep{Done12,Jin12} for thin-disk accretion. In OXAF, the key parameter is $E_{\rm peak}$, the peak energy of the thermal disk emission. For our comparison, we use models with $\log (E_{\rm peak}/\rm keV)=-1.0$, which best reproduce most Seyfert spectra in \citet{Zhu23}.  We adopt the radiative shock models from \citet{Sutherland17}, including both the ``pure shock'' model, dominated by post-shock emission with precursor emission contributing less than $10$\%, and the fast shock'' model, which includes a significant contribution from photoionized precursor emission. For the ``fast shock'' model, the emission from the shocked and the precursor (photoionized pre-shock) regions is summed assuming a fixed ratio of H$\beta$ luminosities of $L_{\mathrm{H}\beta,\mathrm{precursor}} : L_{\mathrm{H}\beta,\mathrm{shock}}=0.5:0.5$ \citep[for details, see][]{Zhu25}. 

{ The AGN photoionization,  ``fast'' shock and ``pure'' shock model grids are shown in Fig.~\ref{fig:emodel} left panel. This comparison shows that: (1) image pixels with SLI $>0.45$ are in general consistent with fast-shock excitation, as well as AGN photoionization. Of these, (2) Pixels with SLI$ >0.45$ and $\log$([O~III]/H$\beta$)$>1.25$ require the fast-shock excitation. (3) Pixels with $0.45>$SLI$>0.25$ can be due to both photoionization and shocks, but not fast shocks. (4) Pixels with SLI $< 0.25$ can only be explained with photoionization models. The right panel of Fig.~\ref{fig:emodel} shows the fast-shock and photoionization model grids together with the VO diagram for ESO~137-G034. { All the points with SLI$>0.45$ are colored. Photoionization can explain most of them, but not those with $\log$([O~III]/H$\beta$)$>1.25$ (green diamonds). These points require shocks with velocities of $v\sim1000$~km~s$^{-1}$. Points with SLI between $0.25$ and $0.45$ can be  explained with either photoionization or ``pure'' shock models. Points with SLI $<0.25$ are only likely to arise from photoionization.

In conclusion, with the SLI, we introduce a method to easily identify regions of different excitation that can then be compared with models and multiwavelength observations (see Section 4).

 \section{Discussion}\label{sec:discussion}

 \subsection{Source of excitation in bi-cones}
\begin{figure*}[thp!]
    \centering
    \includegraphics[width=0.99\textwidth]{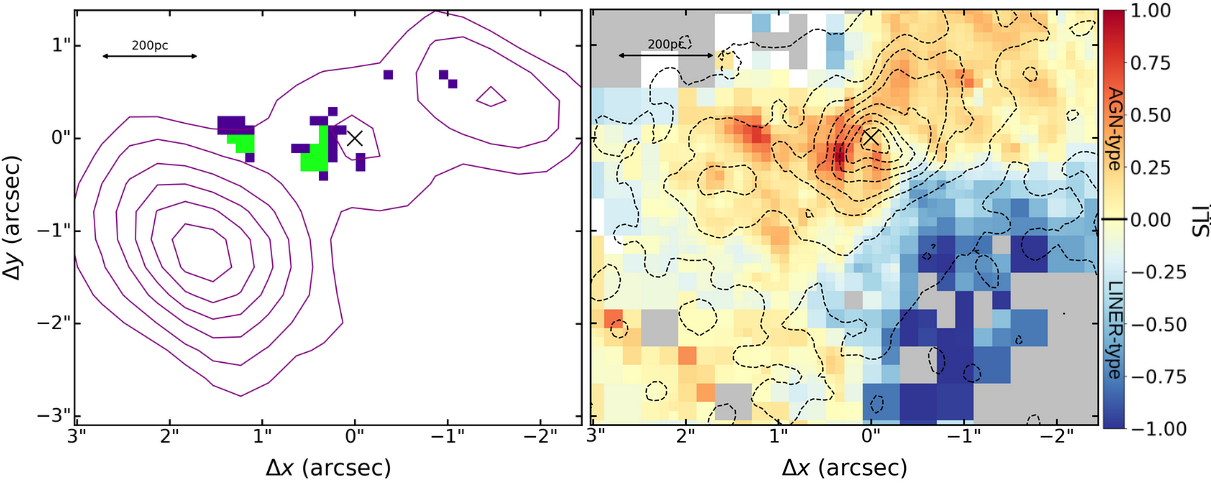}
 \caption{ESO137-G034 maps {illustrating the high-excitation region within the SE cone. Left panel:  purple contours {indicate} the $3$\,cm radio emission \citep[][]{Morganti99}. Here only points with SLI$>0.45$ {  are marked in color, with colors defined as in Fig.~\ref{fig:emodel}}. Right panel: { VO}  {map} of ESO137-G034 with colors defined as in Fig.\,\ref{fig:bpt_sli} with the black dashed contours {tracing} soft ($0.3-1.5$\,keV) X-ray emission (Paper\,I).  The black ``X'' marks the position of the nucleus. }}
    \label{fig:xradio} 
\end{figure*}

 Seyfert-type emission in { BPT { and  VO}  diagrams} is {most commonly} associated with AGN photo-ionization, though {it} can also be produced by fast shocks with a photo-ionizing precursor \citep[][]{Allen08}. {Shock-induced ionization is expected to produce elevated values of $\log$([O~III]/H$\beta$). For instance, fast shocks with velocities $\sim 500–1000$\,km\,s$^{-1}$ can yield $\log$([O~III]/H$\beta$)$\ge1$ \citep[][]{Kewley19}. Exact positions of the AGN and shock excited gas on the { VO}  diagram depend however on the gas metallicity, the strength of the magnetic field and the AGN radiation field \citep[][]{Zhu23}.}

The ionization cones of ESO\,137-G034 are predominantly classified as Seyfert-type{, however only moderately: approximately $90\%$  of the pixels with SLI $>0$ have SLI$<0.45$},  { and their emission is well reproduced by AGN photoionization. }}  { The remaining points with SLI$>0.45$, consistent with fast-shock excitation are mainly present in the SE cone.}

\subsubsection{SE cone}

\begin{figure}[thp!]
    \centering
    \includegraphics[width=0.5\textwidth]{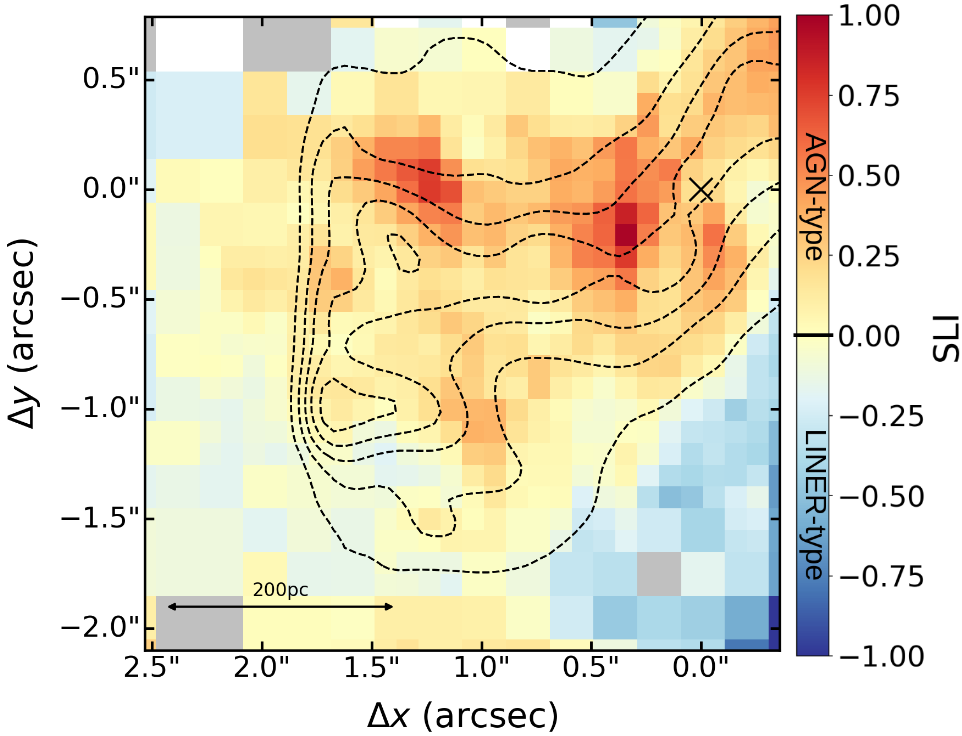}
 \caption{{ VO}-SLI {map} of ESO137-G034 with colors defined as in Fig.\,\ref{fig:bpt_sli}, {illustrating the high-excitation region within the SE cone in different energy ranges. The black dashed contours {trace} [Fe~II] emission.  The black ``X'' marks the position of the nucleus. }}
    \label{fig:feii}
\end{figure}

In the left panel of Fig.~\ref{fig:xradio}}} we map SLI$>0.45$ points positions and over plot $3$\,cm radio contours from \textit{Australia Telescope Compact Array} \citep[ATCA, half power beam width $=1^{\prime\prime}$,][]{Morganti99}. The right panel shows the full SLI map with X-ray ($0.3$-$1.5$~keV) contours and box for reference. {We assumed that the middle maximum of the radio emission corresponds to the source nucleus, defined by the maximum of the [O~III] continuum, indicated by a cross in the Fig.\,\ref{fig:dust}, \ref{fig:bpt} and \ref{fig:bpt_sli}. Contour levels start at the $3\sigma$ level, where $\sigma$ is the standard deviation of the background-dominated region, and increase by a factor of $\sqrt2$ }. 

{ The shock origin induced excitation of SLI$>0.45$ regions is supported by multiwavelength data. These points are clustered along the edge of the radio lobe in the SE cone and }and lie in a larger region of enhanced soft X-ray emission (Paper~I). {Their presence and position on the { VO}  diagram is not affected by the extinction correction (see Fig.\,\ref{fig:extinction}).}In Paper~I, we argue that the X-ray emission in this region (the inner part of the SE cone) is dominated by thermal emission from shock-heated gas, with inferred shock velocities of $\sim800$–$1000$\,km/s. The presence of high velocity shocks at the radio lobe edges have been reported previously by \cite{Croston09} {in Centaurus~A.}

{ { The picture of a significant contribution of the shocks in the SE cone is} supported by the map of the $1.64\mu$m [Fe~II] line in Fig.\,\ref{fig:feii} \citep[obtained as part of a project observing local luminous AGNs, ][]{Davies15}; it is centered by matching the non-stellar continuum peak extracted from the datacube to the location of the AGN as indicated by the black ``X''. This line is enhanced in $\gtrsim100$\,km~s$^{-1}$ shocks which lead to grain destruction \citep[][]{Greenhouse91}, and so is prominent both in supernova remnants and the edges of AGN ionisation cones. The [Fe~II] map shows a ridge of strong emission close to the regions with the highest SLI, which we argue above are indicative of fast shocks. Enhanced [Fe~II] also occurs near the peak of the SE radio lobe, and terminates abruptly at the SLI$\sim0$ transition where the line emission changes from being Seyfert like to LINER like.
{ The LINER region (negative SLI values) are generally located close to the Seyfert–LINER transition line, with SLI values between $\sim-0.1$ and $0$. The morphology of LINER points in the SE cone is patchy.}  These are all indicators that the [Fe~II] is tracing shocks at the edge of the ionisation cone that are also associated with the radio jet impinging on ambient ISM.}

\subsubsection{NW cone}\label{sec:NW}

The Seyfert-like points in the NW cone show less extreme SLI values compared to those in the SE cone{: $97$\% of Seyfert-type pixels have SLI$<0.45$, consistent with AGN photoionization}. { The exception is the concentration of elevated SLI ($>0.45$) values at the eastern edge of the NW cone. However, this high-SLI region coincides with the dust lane (see Fig.~\ref{fig:dust} for comparison). The SLI values characterizing that region decreased in the reddening-corrected image { to SLI$\sim0.2$} (Fig.~\ref{fig:extinction}), and therefore we conclude that high SLI values originate from extinction.}
The radio jet in the NW cone direction is weaker,{ and there is no corresponding enhancement in soft X-ray emission, unlike what is observed in the SE cone. }This is consistent with the SLI values indicating that the NW cone is {primarily driven} by AGN photo-ionization.

\begin{figure*}[t!]
    \centering
    \includegraphics[width=0.53\textwidth]{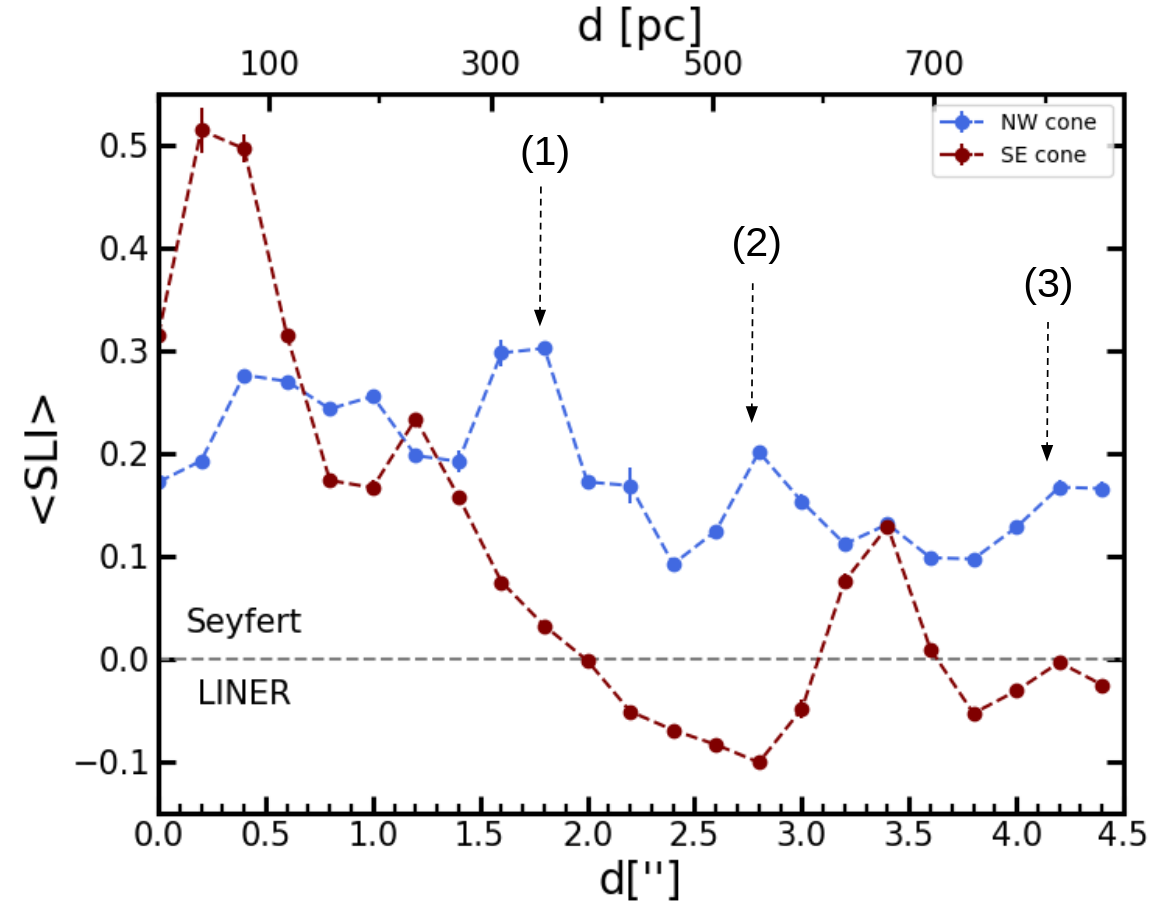}
    \includegraphics[width=0.46\textwidth]{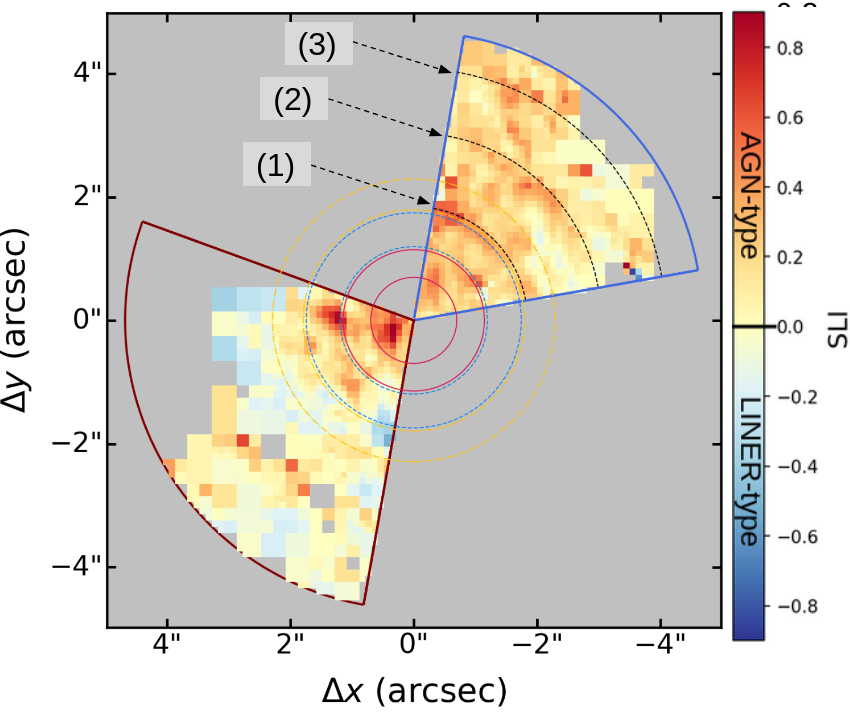}
 \caption{{ Radial profile of the SLI for the bi-cone regions of ESO~137-G034. Left panel: radial profile of the average SLI value for the points in the SE (red) and NW (blue) ionization cones. Light blue/red lines and shaded regions correspond to the radial profile and its errors for the reddening-corrected data. Arrows mark three maxima in NW SLI profile. Right panel: SLI map with marked radial SLI profile extraction regions in the SE (red) and NW (blue) ionization cones. We mark arcs representing the positions of the SLI maxima in with dashed black lines. { Three rings mark the projected radial bins $120$--$240$~pc (solid red), $240$--$360$~pc (dashed blue), and $360$--$480$~pc (dot-dashed yellow), from which the azimuthal profiles shown in Fig.~\ref{fig:profiles} were extracted.}}}
    \label{fig:profile_rad}
\end{figure*}

{ The radial profile of the SLI values (Fig.~\ref{fig:profile_rad}), shows that the average SLI decreases radially, with values ranging from $\sim0.25$ at $r=0$--$500$~pc to $\sim0.15$ at $R=500$--$750$~pc. This radial decrease is consistent with a radial decrease of the ionization parameters with the distance from the AGN. Within the decreasing SLI trend we notice localized peaks at $\sim1.8''$, $\sim2.9''$ and $\sim4.0''$, corresponding to physical radii of $\sim350$\,pc, $\sim550$\,pc and $\sim750$\,pc.  }
{ These correspond to the arcs of {enhanced excitation visible} in the right panel of Fig.\,\ref{fig:profile_rad}. In the { VO}  diagram, these regions have enhanced SLI $\sim 0.2-0.3$, (see left panel of Fig.~\ref{fig:profile_rad}). The inner ``ridge'', with SLI$\sim0.3$, could be consistent with either photoionization or pure shock models, but the other two ridges have SLI values typical of photoionization (see left panel of Fig.~\ref{fig:profile_rad}).   Therefore, the ``ridges'' most likely originate from episodes of enhanced nuclear continuum emission. In this scenario, the NW cone would have a constant or smoothly decreasing density, and the enhanced SLI would result from a temporarily higher ionizing flux. The presence of these features is independent of the reddening correction.}

Assuming that the ridges arise from { successive activity cycles, they imply timescales of $\sim(1-2)\times10^3$\,yrs based on light-travel time, or $\sim(5-10)\times10^3$\,yrs assuming $0.2$\,c a typical radio jet hot spot propagation velocity at these distances \citep[][]{An12}.} These time scales are comparable to the length of activity cycles driven by the radiation pressure instability within the accretion disk \citep{Czerny09}. For typical outflow velocities of $\sim1000$\,km\,s$^{-1}$ \citep{Crenshaw03}, the implied activity timescales would be $\sim (3-7)\times10^5$\,yrs.  The widths of the enhanced SLI arcs are $\lesssim0.3^{\prime\prime}$, corresponding to $\lesssim100$\,yr, suggesting  short episodes of enhanced nuclear activity. Short activity episodes have been detected in AGNs \citep{Soldi08} and in the SgrA* black hole at the center of the Milky Way \citep{Do19}. Alternatively, ridges may reflect the irregularities in the ISM structure, { but their radial character suggests link to the AGN activity.}

\subsection{Comparison between SE and NW Cones}

The difference between the SE and NW cones is well illustrated by the radial profiles of the SLI value (Fig.\,\ref{fig:profile_rad}). The NW cone (blue line) exhibits relatively stable values with three distinct maxima, { as discussed in Sec.~\ref{sec:NW},} while the SE cone shows significantly greater variation. {This is due to the presence of both inter-cone LINER regions { (negative SLI)} and Seyfert-like regions with very high relative values of SLI.} 

\begin{figure}[thp!]
    \centering
    \includegraphics[width=0.49\textwidth]{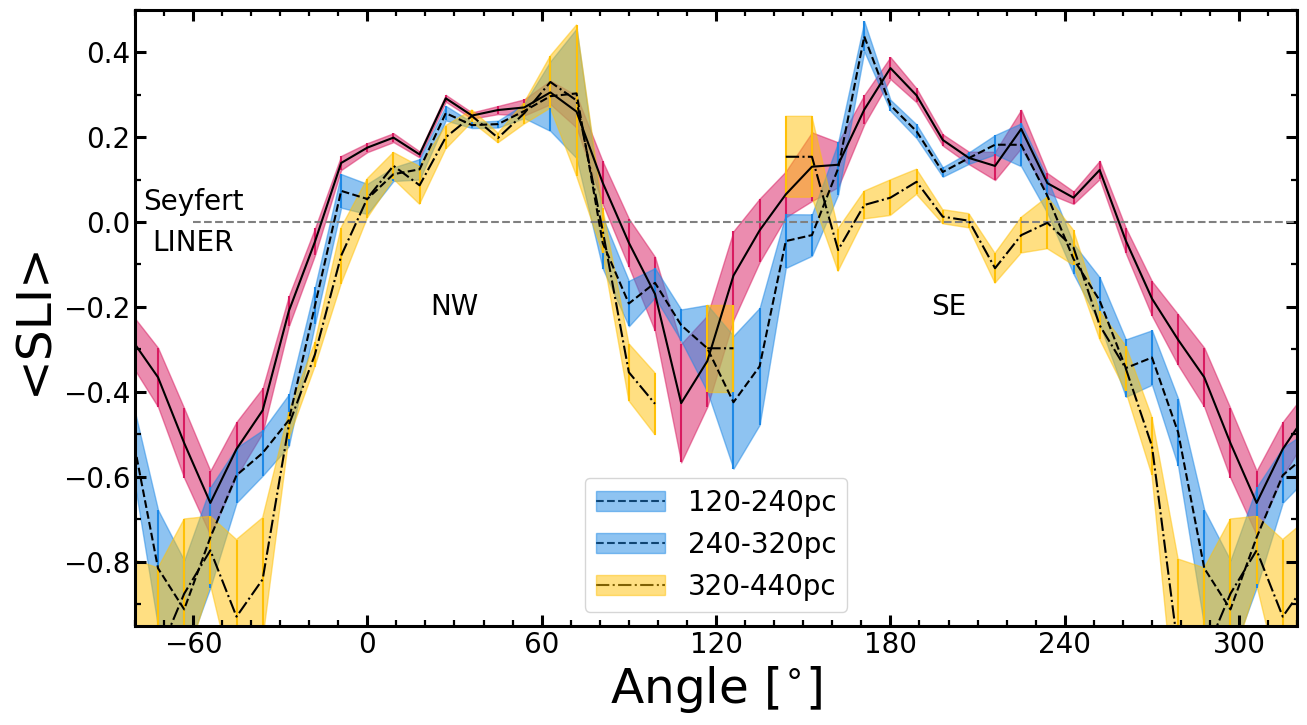}
 \caption{{ Azimuthal profiles of the mean SLI in three projected radial bins: $120$--$240$~pc (solid red), $240$--$360$~pc (dashed blue), and $360$--$480$~pc (dot-dashed yellow). The extraction regions are marked in the right panel of Fig.~\ref{fig:profile_rad}. Shaded regions indicate the uncertainties.} }
    \label{fig:profiles}
\end{figure}

{ This difference is further highlighted by the azimuthal profiles of the mean SLI. In Fig.\,\ref{fig:profiles}, we show the azimuthal variation of the mean SLI in three radial bins: $120$--$240$~pc, $240$--$360$~pc, and $360$--$480$~pc.  The azimuthal profiles within the NW cone ($-10^{\circ}$ to $80^{\circ}$) are self-similar across radii, whereas the SE cone shows radial dependence.}  Differences in emission properties are also observed in the X-rays (Paper~I), with the SE cone showing significant variation in the azimuthal profiles of extended emission between the soft ($0.3$–$1.5$\,keV) and hard ($1.5$–$3.0$\,keV) energy bands. As shown in Fig.\,\ref{fig:xradio}, radio emission is present in both directions, but {but is stronger and exhibits a well-defined radio-lobe morphology in the SE cone.}

These multi-wavelength morphological differences suggest that the properties of the ISM into which the jet or outflow propagates differ between the NW and SE directions, with the SE cone likely encountering a more clumpy medium.

\subsection{LINER Cocoon of the NW cone}

{ Observationally, in { VO}  maps, extended Seyfert excitation regions appear to be surrounded by LINER regions \citep[dubbed LINER cocoons; e.g.,][]{Ma21,Maksym16,Cresci15}. In general, the LINER-type emission in an AGN context is associated either with photoionization by a low-luminosity or obscured AGN, or with lower-velocity shocks without a photoionizing precursor \citep[''pure shocks``, ][]{Dopita95,Ho96,Kraemer08,Halpern83}. The origin of the LINER cocoons, in particular, is either linked to obscured AGN radiation or to shocks arising from interactions between AGN winds and the surrounding ISM \citep[][]{Ma21}. }

{
\citet{Ma21} showed that some LINER cocoons occupy intermediate positions between Seyfert and LINER classifications on the { VO}  diagram. Here, Fig.~\ref{fig:profiles} demonstrates that the SLI transition between Seyfert- and LINER-type { across different radii in the NW cone  }is smooth and monotonic, with values SLI $\sim 0$. The smooth transition suggests a layered ionization: the cocoon is approximately $\sim100$–$200$~pc thick and radially extends at least up to $\sim0.5$~kpc from the nucleus, with no signs of radial dependence. The SLI { azimuthal} profiles of the LINER cocoon are continuous, with no indication of strong patchiness. 

{  An exception is the elevated SLI values at $\sim80^{\circ}$ within the NW cone,}coinciding with the dust lane and likely originating from the extinction.}

{ The similarity of the NW cone SLI profiles across radii indicates that the cone geometry is preserved and remains largely constant with radius, aside from the SLI ridges seen in the radial profile (Fig.~\ref{fig:profile_rad}). }

Complex kinematic profiles with outflow signatures are common in the ionized ISM of Seyfert galaxies \citep{Fischer13}. {Therefore, the LINER cocoon around the NW cone may arise from lateral expansion of the outflowing wind. Another possible source of excitation is photoionization by an AGN spectral energy distribution that is affected by extinction from a partially ionized medium, i.e. a ''warm absorber`` (WA) at the base of the cone near the AGN \citep[][]{Halpern83}. Both mechanisms may operate simultaneously.}

In the case of an extinction scenario, the lack of a radial SLI gradient in the cocoon of the NW cone (Fig.\,\ref{fig:bpt_sli}) suggests that the absorber is located closer to the nucleus than the spatial scales resolved with SLI (i.e., $\ll100$\,pc). If we assume that the absorbing gas in ESO\,137-G034 is analogous to the WA seen in many Seyfert type 1 galaxies, this conclusion is consistent with estimates and upper limits placed on WA–nucleus distances in some other sources, e.g., NGC~985 \citep[$<20$\,pc][]{Krongold05}, NGC~3783 \citep[$<50$\,pc][]{Gabel05}, and NGC~4051 \citep[$<1$\,pc][]{Krongold07}.

\subsection{ Cross-cone excitation}
In the cross-cone regions, we {observe a mix of} H~II-type and LINER-type (SLI$<0$) points. The position of LINER-type points on the { VO}  map appears independent of the distance from the Seyfert-type region or the nucleus, as indicated by the plateau in the profiles for cross-sections at distances $<0.1$\,kpc in the case of both cones (Fig.~\ref{fig:profiles}).

\section{Summary and Conclusion}\label{sec:summary}

In this work, we used narrow-band HST imaging to construct the spatially resolved { VO } diagram and map of CT AGN ESO\,137-G034. Additionally, we  introduced a Seyfert/LINER Index (SLI), defined as the perpendicular distance of each point from the Seyfert/LINER division line on the { VO}  diagram.{ By introducing SLI, we were able to visualize fine-scale excitation variations in the ISM. That helped us differentiate shock-excited gas from other AGN-driven Seyfert-like emission, and identify the radial ridges in the NW cone ionization state}. Our findings can be summarized as follows:

\begin{itemize}
    \item The emission in the [O~III], [S~II], H$\alpha$ and H$\beta$ lines traces a bi-cone morphology, with an elongation in the North-West and South-East directions. { The optical emission} traces closely the X-ray diffuse emission seen in X-rays by {\it Chandra}. 
    \item The majority of the emission within the bi-cones is Seyfert-type and is characterized by moderate SLI values ($90$\% of the Seyfert pixels have SLI$<0.4$).Based on theoretical models for most of these points, the { likely} source of excitation is AGN photoionization. 
    \item { All regions with SLI $>0.45$ are consistent with fast-shock excitation. For some, shock-induced excitation is the only viable origin, while for others the emission can be also explained by photoionization. }
    \item The SLI$>0.45$ pixels are clustered in the SE cone inner region ($\lesssim200$\,pc from the nucleus).  These points overlap with a region of soft X-ray emission excess and with the inner edge of the radio lobe. Their position on the { VO}  diagram is consistent with excitation by fast shocks ($\gtrsim1000$\,km\,s$^{-1}$). The remaining Seyfert-classified pixels in the SE cone, which have lower SLI ($<0.4$) values, are consistent with AGN photoionization, but SE cone shows widespread SLI inhomogeneities.
    \item The NW ionization cone displays a relatively smooth excitation profile, dominated by AGN photoionization. Three higher SLI values ($\sim0.2$--$0.3$) arcs located at projected radii of $\sim350$, $\sim550$, and $\sim750$\,pc are found in the NW cone. These ridges may reflect past AGN activity cycles, with inferred timescales of $\sim(1$--$2)\times10^3$ years assuming light-travel time, up to $\sim(5$--$10)\times10^3$ years for a jet propagation velocity of $0.2c$, or $(3-7)\times10^5$\,yrs for typical [OIII] outflow velocities. Alternatively, these features may be linked to nonuniform profile of the ISM. { Overall average SLI value decreases by $\sim0.1$ radially.}
    \item A LINER-type cocoon envelops both ionization cones, most prominently in the NW direction, where it appears as a smooth and spatially continuous structure. The LINER cocoons are $\sim 100$--$200$\,pc thick and extends out to at least $0.5$\,kpc from the nucleus. The gradual SLI transition across cone boundaries suggests a layered ionization structure, shaped by AGN winds interacting with the ISM, obscured AGN radiation, lower-velocity shocks, or a combination of these. 
\end{itemize}

In conclusion, the use of the SLI uncovers new structure within the AGN and LINER zones of the { VO}  diagram { and allows to spatially distinguish different excitation mechanisms. { With the caveat that high signal to noise data is needed to distinguish different ISM excitation structures, our method has the potential of uncovering spatial complexities of the AGN feedback. Firmly establishing he general applicability of the SLI will require the systematic analysis of larger samples of AGNs, which we propose to undertake in the future. This} technique requires spatially resolved maps of fluxes in [S~II], H$\beta$, H$\alpha$ and [O~III]. These can be obtained by HST, or with worse spatial resolution but better energy resolution, by the Multi Unit Spectroscopic Explorer (MUSE). }

In the case of ESO137-G034 the {\it Chandra} data does not have sufficient signal-to-noise to see if the SLI-discovered features produce the complex photoionized and thermal X-ray spectra. A larger area sub-arcsecond  X-ray spectral imaging mission, preferably with higher resolving power \citep[Lynx Consortium, ][]{Gaskin19} could explore the SLI connections to the multi-component outflows of AGNs.

\begin{acknowledgments}
    This work was supported by grant  HST-16841.002-A and partially by NASA contract  NAS8-03060 (CXC). It was performed in part at  Aspen Center for Physics, which is supported by National Science Foundation grant PHY-2210452.
\end{acknowledgments}
\appendix

\section{{ VO}  diagrams for different H$\alpha$-[N~II] ratios}\label{sec:app}
\begin{figure}[thp!]
    \centering
    \includegraphics[width=0.99\textwidth]{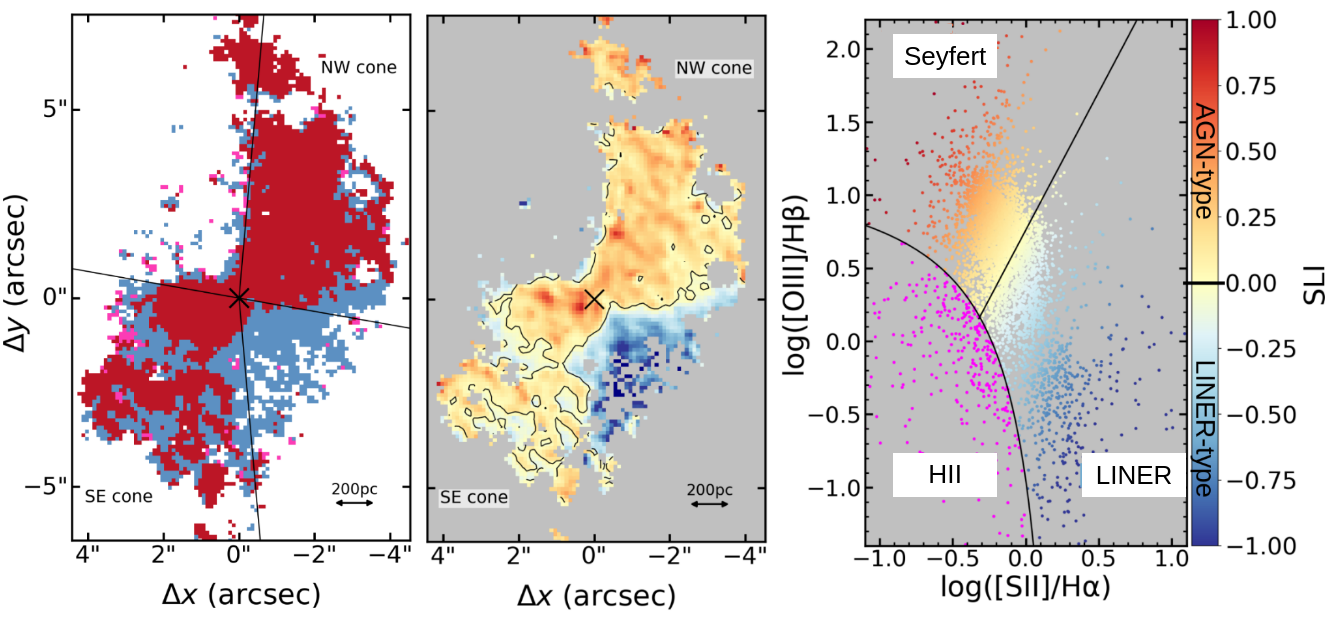}
    \includegraphics[width=0.98\textwidth]{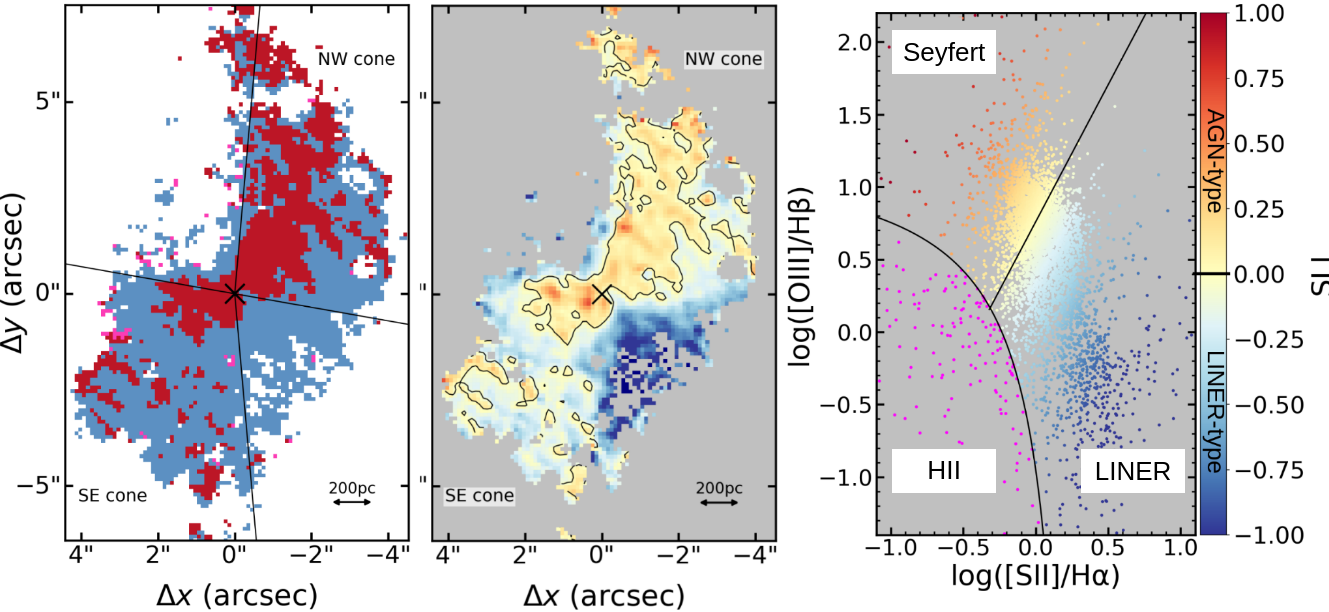}
    \caption{Spatially resolved { VO}  maps and diagram of ESO137-G034 with colors as in Fig.\,\ref{fig:bpt} and Fig.\,\ref{fig:bpt_sli}, for assumed $35\%$ (upper row) and $55\%$ (lower-row) contribution of the H$\alpha$ emission to the total flux in FR680P15 filter. }
    \label{fig:Ha}
\end{figure}

We examined the robustness of our results by varying the assumed H$\alpha$/(H$\alpha$+[~NII]) ratio of $0.45$.  Fig.\,\ref{fig:Ha} shows the { VO}  maps and diagrams obtained for $0.35$ and $0.55$ values. {While these variations lead to the reclassification of some individual pixels, the overall spatial distribution and morphological features of the excitation regions remain consistent. The main conclusions of our analysis are therefore unaffected by the specific choice of H$\alpha$-[N~II] decomposition.}

\newpage

\section{Extinction correction}\label{sec:app2}

\begin{figure}[thp!]
    \centering
    \includegraphics[width=0.49\textwidth]{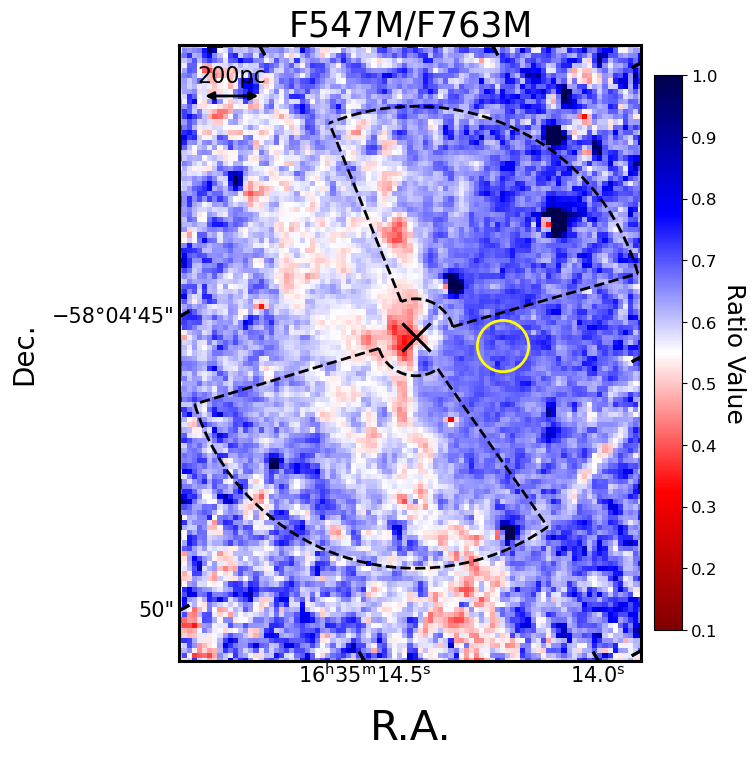}
    \includegraphics[width=0.49\textwidth]{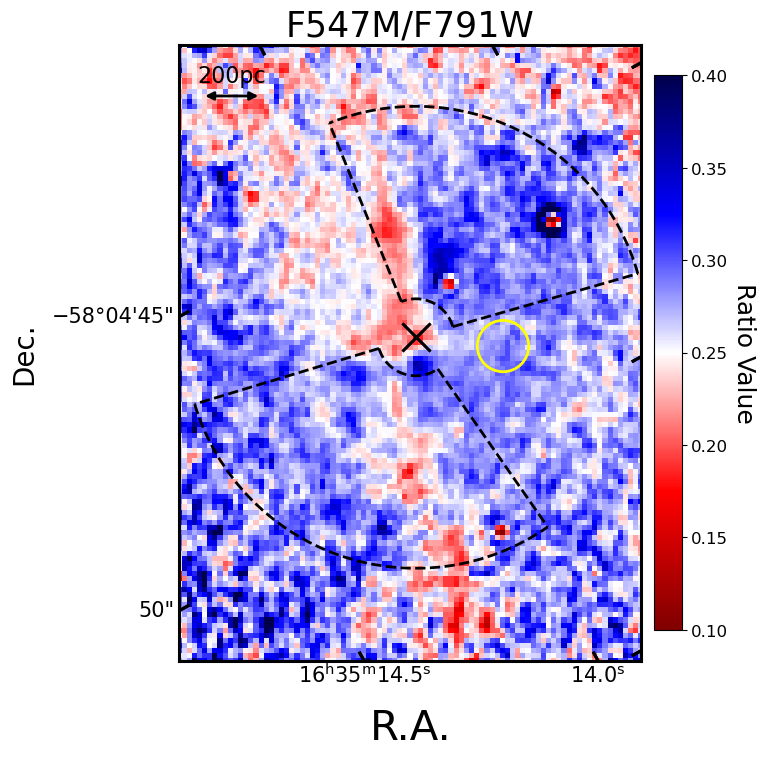}
 \caption{Blue to red continuum ratio maps for two different sets of filters: F\,547M to F\,791W (upper panel) and F\,547M to F\,763M (lower panel). Red regions indicate areas most affected by dust. {Dashed black lines mark ionization cones defined by diffuse X-ray emission (Paper~I). The black ``X'' marks position of the nucleus, defined as the centroid of the [O~III] continuum emission. { The yellow circle marks the low-dust region used for the extinction correction.  }} }
    \label{fig:dust}
\end{figure}

\begin{figure}[thp!]
    \centering
    \includegraphics[width=0.99\textwidth]{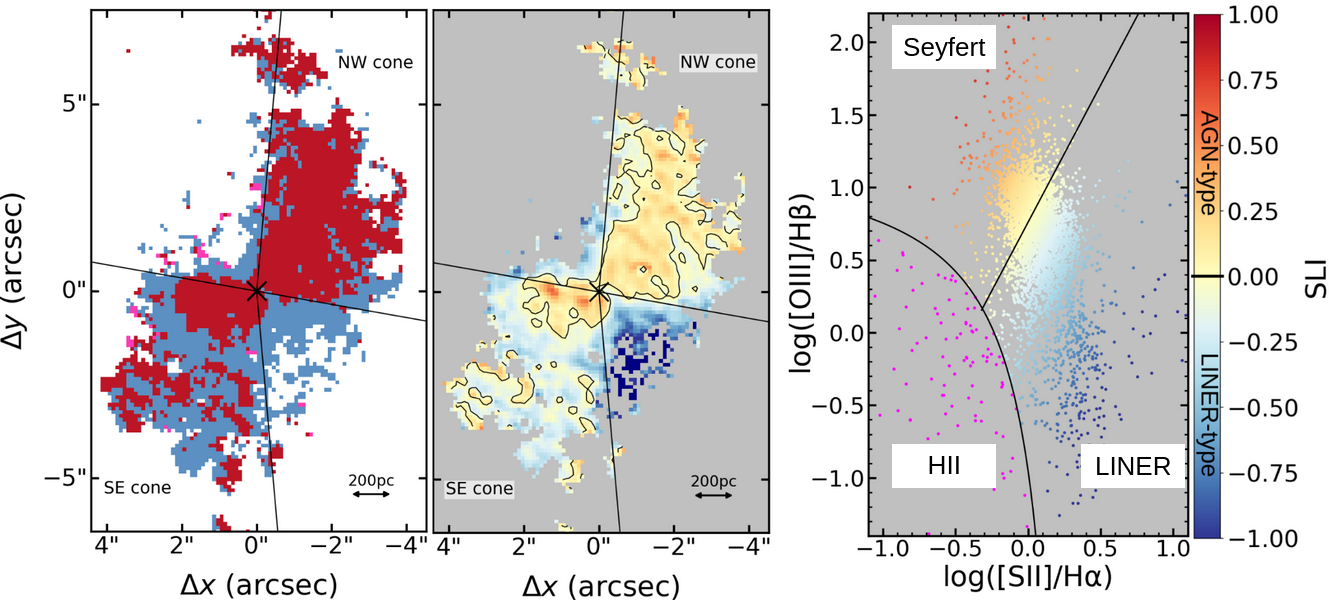}
 \caption{Spatially resolved { VO}  maps and diagram of ESO137-G034 with colors as in Fig.\,\ref{fig:bpt} and Fig.\,\ref{fig:bpt_sli}, for extinction corrected data. }
    \label{fig:extinction}
\end{figure}

{To assess the effects of dust extinction on the fluxes measured in the narrow and continuum filters, we performed a reddening correction following the approach of \citet{Maksym21}. Based on the blue-to-red continuum maps derived for two filter pairs, F547M to F791W (WFPC2) and F547M to F763M (WFC3) (Fig.~\ref{fig:dust}), we identified the regions least affected by dust and adopted these as representative of the intrinsic stellar population color. Using this as a reference, we applied a wavelength-dependent, pixel-by-pixel reddening correction to all filters, following the \citet{Calzetti00} extinction law \citep[see Appendix \ref{app:errors} here and Appendix A in][]{Trindade25}. We calculated the color excess separately for the WFPC~2 and WFC~3 observations, and then averaged value. Corrections were applied for  both the narrow-band filter and the continuum observations. Using extinction-corrected expositions, we constructed the continuum-subtracted narrow-line emission images of ESO~
137-G034 in [O~III], [S~II], H$\beta$, and H$\alpha$, and used them to create the { VO}  and SLI maps in Fig.~\ref{fig:extinction}. The left panel shows the { VO}  map, the middle panel shows the map of SLI values, and the right panel shows the { VO}  diagram with points color-coded to reflect the SLI value for a given point. The extinction corrections affect only the source regions covered by the dust lane (red regions of Fig.~\ref{fig:dust}). The main results of our analysis are not affected by the reddening correction.}

\section{Error propagation}\label{app:errors}
{ The resulting SLI uncertainty is derived using the standard error propagation formula for uncorrelated variables:}
\begin{equation}
\sigma_{f} = \sqrt{\sum_{i}\big(\sigma_{x_i}\partial_{x_i} f\big)^2},
\end{equation}
{ where $\sigma_{x_i}$ is the uncertainty associated with variable $x_i$, $f$ is a function of the variables $x_i$, and $\sigma_{f}$ is the corresponding propagated uncertainty.

We adopted the standard deviation of the pixel counts in a background-dominated region, as the uncertainty for each individual filter observation \citep[][]{Ma21}. Then we propagated these errors  through the consecutive steps of the data analysis on a pixel-by-pixel basis, starting from the continuum subtraction for [O~III], [S~II], H$\alpha$, and H$\beta$ emission lines. The narrow-band flux is defined as: }
\begin{equation}
    f_l = \alpha_n (n- n_{bkg}) - \alpha_c\beta(c- c_{bkg}),
\end{equation}
{    with $\alpha$ denoting the product of {\tt PHOTFLAM} and {\tt PHOTBW}, header keywords converting the image units to erg s$^{-1}$cm$^{-2}$, for the narrow and continuum filters. $\beta$ is defined as the ratio of narrow to continuum bandwidths, $f_l$ is the resulting flux, and $n$ and $c$ are the count rates in the narrow and wide filters, respectively, with the associated $ n_{bkg}$ and $c_{bkg}$ mean background levels. $f_l$ was calculated for all four optical emission lines. The $f_l$ error, $\sigma_{f_l}$, can be expressed as:}
\begin{equation}
    \sigma_{f_{l}}=\sqrt{\alpha_n^2\sigma_n^2+\alpha_c^2\beta^2\sigma^2_c},
\end{equation}
{ with $\sigma_n$ and $\sigma_c$ being the uncertainty for the narrow and wide filter observations. }

{ Next, we calculated for each pixel the logarithm of the [O~III] to H$\beta$ flux ratio, $Y=\log(f_{O3}/f_{H\beta})$, and the logarithm of the [S~II] to H$\alpha$ flux ratio, $ X = \log(f_{S2}/f_{H\alpha})$. The associated errors can be expressed as: } 
\begin{equation}
    \sigma_Y=\frac{1}{f_{O3}\ln10 }\sqrt{\sigma^2_{f_{O3}}+f_{O3}^{-2}f^2_{H\beta}\sigma^2_{f_{H\beta}}}
\end{equation}
{ and similarly: }
\begin{equation}
    \sigma_X=\frac{1}{f_{S2}\ln10 }\sqrt{\sigma^2_{f_{S2}}+f_{S2}^{-2}f^2_{H\alpha}\sigma^2_{f_{H\alpha}}}
\end{equation}
{ In the final step, we calculate the SLI and the associated errors. SLI is defined as the perpendicular distance of the point to the Seyfert/LINER division line in the { VO } diagram; therefore it can be expressed as:}
\begin{equation}
SLI = \frac{1.89X - Y + 0.76}{2.14},
\end{equation}
{ with the associated error:}
\begin{equation}
\sigma_{SLI} = \sqrt{0.22\sigma_{Y}^2 + 0.68\sigma_X^2}.
\end{equation}

{ If an extinction correction is applied, the narrow-band and continuum count-rate uncertainties have an additional component. Following the \cite{Maksym16} and \cite{Trindade25} approach, we introduce the reddening correction as:}
\begin{equation}
    x_{ob} = x_{int}\times10^{-0.4A\lambda},
\end{equation}
{ with $x_{obs}$ and $x_{int}$ denoting the observed and intrinsic emission, and $A_\lambda$ the extinction at the wavelength $\lambda$. $A_\lambda$ can be calculated from the formula: }
\begin{equation}
    A_{\lambda}=k_{\lambda}E(B-V) =  \frac{k_{\lambda}}{-0.4(k_B-k_R)}\log\bigg(\frac{c_B}{c_RC}\bigg),
\end{equation} 
{ with the reddening curve $k_{\lambda}$ being defined for each {\it HST} filter, and the color excess, $E(B-V)$, calculated for each pixel based on}:
\begin{equation}
    E(B-V) = \frac{1}{-0.4(k_B-k_R)}\log\bigg(\frac{c_B}{c_RC}\bigg).
\end{equation}
{ The B and R indexes refer to blue and red continuum respectively in the ratio of continuum count rates ${c_B}$/${c_R}$. $C$ is the value of the continuum ratio in a region not affected by dust, assumed to be the unabsorbed color ratio (yellow circle in Fig.~\ref{fig:dust}). The error of the reddening correction is given by }
\begin{equation}
    \sigma_{x_{int}} = 10^{0.4 A_{\lambda}}\sqrt{\sigma_{x^2_{obs}}+0.16[\ln(10)]^2x^2_{obs}\sigma_{_{\lambda}}^2}.
\end{equation}
{ With the error associated with $A_{\lambda}$, defined as:}
\begin{equation}
    \sigma_{A_{\lambda}} = \frac{k_{\lambda}}{\ln10|k_B - k_R|}\sqrt{\sigma_{c_B}^{2}c_B^{-2} + \sigma_{c_R}^{2}c_R^{-2}},
\end{equation}
{ where $k_{\lambda}$ is the value of the reddening curve for the filter for which the correction is calculated.}

\bibliography{sample631}{}
\bibliographystyle{aasjournal}

\end{document}